\documentstyle[preprint,prabib,aps]{revtex} 
\begin{document}
\draft
\title{Einstein-Cartan-Heisenberg Theory of Gravity \\
with Dynamical Torsion}
\author{V. Dzhunushaliev 
\thanks{E-Mail Address : bars@krsu.edu.kg}} 
\address{Universit{\"a}t Potsdam, Institut f{\"u}r Mathematik,
14469, Potsdam, Germany \\
and Theor. Phys. Dept. KSNU, 720024, Bishkek, Kyrgyzstan}
\author{D. Singleton 
\thanks{E-Mail Address : das3y@maxwell.phys.csufresno.edu}} 
\address{Dept. of Phys. CSU Fresno, 2345 East San Ramon Ave.
M/S 37 Fresno, CA 93740-8031, USA} 

\date{\today}
\maketitle

\begin{abstract}
On the basis of an algebraic relation between torsion 
and a classical spinor field a new interpretation 
of Einstein-Cartan gravity interacting with a classical 
spinor field is proposed. In this approach the spinor 
field becomes an auxiliary field and the dynamical 
equation for this field (the Heisenberg equation)  is a 
dynamical, gravitational equation for torsion. The simplest 
version of this theory is examined where the metric 
degrees of freedom are frozen and only torsion plays a 
role. A spherically symmetric solution of this 
theory is examined. 
This solution can be interpreted, in the spirit of Wheeler's 
ideas of ``charge without charge'' and ``mass without mass'', 
as a geometrical model for an uncharged 
particle with spin (``spin without spin''). 
\end{abstract}

\section{Introduction} 

The simplest generalization of general relativity 
is Einstein-Cartan (EC) gravitational theory in which 
torsion is included. Remarkably this theory 
is a gauge theory of gravity similar in spirit to  
Yang-Mills gauge theories. The difference being that the
gauge group of the former is the Poincare group, while
in the latter case one has gauge groups like SU(3) or SU(2).
One peculiarity of Einstein-Cartan theory is that the torsion
does not have a dynamical term in the Lagrangian and is therefore 
a nonpropagating field. This fact stimulated 
research for a new dynamical gauge theory of gravity. 
Reviews of such attempts can be found in  Refs. \cite{hehl1}, 
\cite{hehl2}. Also an overview of the geometrodynamics program applied to
Yang-Mills theories and gravitational theories with torsion
can be found in Ref. \cite{ewm1}.
The Lagrangians of these theories contain terms which are
quadratic in the curvature and/or torsion. Thus, in these
theories the price for making the torsion dynamic is that one
must introduce quadratic terms into the Lagrangian. 
\par
In this paper we attempt to make torsion a dynamical
quantity, while leaving the EC Lagrangian unchanged.
This is done by introducing an auxiliary classical spinor
field.  The EC field equations in this case
give an algebraic relation between the torsion and spinor field 
that allow us to interpret the nonlinear 
Heisenberg equation for the spinor field as a dynamical 
equation for torsion \cite{dzh1}. The
reverse connection that is made in this paper between
spin and torsion gives a geometrization of spin. This idea
(that spin should have a geometrical interpretation) goes back to
Wheeler \cite{wh}. Along similar lines a Rainich type
geometrization of spin has been carried out in Ref. \cite{kuc}.

\section{Einstein-Cartan gravity with spinor field}

The Lagrangian for EC gravity with a classical spinor field 
can be written as (in this section we follow the notation of
Ref. \cite{dzh1}): 
\begin{equation}
L=\frac{\hbar c}2\left[ i\overline{\psi }
\gamma ^\mu(\nabla _\mu\psi )-\frac{mc} 
\hbar \overline{\psi }\psi +(Hermitian-conjugate)\right] - 
\frac 1{2k} R,  
\label{1} 
\end{equation} 
here Greek indices $=0,1,2,3$ are 4D spacetime
indices; $\gamma ^\mu$ are Dirac matrices satisfying  
$\{\gamma ^\mu,\gamma ^\nu\}=2g^{\mu\nu}$; $g^{\mu\nu}$ 
is the 4D spacetime metric; $k=8\pi G/c^4$; 
$R = g^{\beta\gamma} R^\alpha_{\beta\alpha\gamma}$ 
is the 4D Ricci scalar of the affine connection 
$\Gamma ^A_{\bullet BC}$ (The definitions of the
various Riemann-Cartan geometrical objects are given
in Appendix A). A modern account of
EC-Dirac theory with torsion is given in Ref. \cite{ewm2}.
In Eq. (\ref{1}) we have included the standard mass parameter, $m$.
Heisenberg's original idea in his studies of
the non-linear Dirac or Heisenberg equation was
that one should set $m=0$, and that
the mass should appear dynamically 
via $M = \int T_{00} d^3 x$ 
($T_{\alpha\beta}$ is energy-momentum tensor for 
the spinor and electromagnetic fields). 
The covariant derivative of the spinor 
field is defined as \cite{sab}, \cite{hehl}: 
\begin{equation} 
\nabla _\mu\psi =\left( \partial _\mu-\frac{1}{4}
\omega _{ab\mu}\gamma ^{[a}\gamma 
^{b]}-\frac 14S_{ab\mu}\gamma ^{[a}\gamma ^{b]}\right) \psi ,
\label{2}
\end{equation}
here $a,b=0,1,2,3$ are vier-bein indices; 
$\gamma ^a = \gamma ^0, \gamma ^1, \gamma ^2, \gamma ^3$ 
are ordinary Dirac
matrices $\{\gamma ^a,\gamma ^b\}=2\eta ^{ab}$; 
$\gamma^\mu = h^{\bullet\mu}_a \gamma^a$ 
($h_{\bullet \mu}^a$ is a vier-bein);  
$\eta ^{ab} = diag\{ 1,-1,-1,-1\}$ 
is the 4D Minkowski metric; [ ] means antisymmetrization. 
The coefficients of the spinor 
connection are defined as follows: 
\begin{equation} 
\omega _{ab\mu} = h_{a\alpha}h_b^{\bullet \nu} 
\left \{ \,^\alpha_{\mu\nu}\right \} + 
h_a^{\bullet \nu}\frac{\partial h_{b\nu}}{\partial x^\mu}    
\label{3} 
\end{equation} 
here $\left \{ \,^\alpha_{\mu\nu}\right \}$ 
are the Christoffel symbols. 
If the torsion is taken as totally antisymmetric then 
varying the torsion, spinor fields and metric 
leads to the following fields equations: 
\begin{eqnarray}
S^{abc} = 4il_{Pl}^2\left (\overline{\psi }
\gamma ^{[a}\gamma ^b\gamma ^{c]}\psi \right ), &&
\label{4} \\
\left( i\gamma ^\mu\partial _\mu-\frac{i}{4}
\omega _{ab\mu}\gamma ^\mu\gamma
^{[a}\gamma ^{b]} + l_{Pl}^2(\overline{\psi }
\gamma ^{[a}\gamma ^b\gamma
^{c]}\psi )\gamma _{[a}\gamma _b\gamma _{c]} - 
\frac{mc}\hbar \right) \psi 
&=&0,  
\label{5} \\
R_{\mu\nu}-\frac {1}{2}g_{\mu\nu}R = 8l_{Pl}^2T_{\mu\nu}^D - 
8 l_{Pl}^4g_{\mu\nu}(\overline{\psi }
\gamma ^{[a}\gamma ^b\gamma ^{c]}\psi )
(\overline{\psi }\gamma _{[a}\gamma
_b\gamma _{c]}\psi ), &&  
\label{6}
\end{eqnarray}
here $S^{abc}$ is the antisymmetric torsion tensor, 
$l_{Pl}=\sqrt{\pi\hbar G/c^3}$ is the Planck length, 
$R_{\mu\nu}$ is the Ricci tensor, $g_{\mu \nu}$
is the metric which is defined in the ordinary way
$g_{\mu\nu}=h_{\bullet \mu}^a h_{a\nu}$. 
The energy-momentum tensor $T_{\mu\nu}^D$ of the Dirac field is:  
\begin{equation} 
T_{\mu\nu}^D = {i\over 4}\left[ \overline{\psi }
\gamma _\mu(\stackrel{\{\}}{\nabla }_\nu\psi
) + \overline{\psi} \gamma _\nu
(\stackrel{\{\}}{\nabla }_\mu\psi ) \right]
+(Hermitian-conjugate),  
\label{7}
\end{equation} 
here $\stackrel{\{\}}{\nabla }_\mu$ 
means the covariant derivative without torsion.

\section{Dynamical torsion from Heisenberg equation. \\
Einstein-Cartan-Heisenberg gravity}

Eq.(\ref{4}) establishes 
an {\it algebraic connection} between the
torsion and the spinor field. Ordinarily the 
right hand side of this equation (the spinor field) is 
interpreted as the source of the left side (the torsion). 
We propose \cite{dzh1} the {\it reverse} interpretation: 
torsion is the source for the classical spinor field. 
In the 4D case the totally antisymmetric torsion can be 
represented as the pseudo-vector 
$S^\mu = \epsilon ^{\mu\beta\gamma\delta}
S_{\beta\gamma\delta }$ 
and has 4 independent degrees of freedom. The
spinor $\psi$ also has 4 components. Hence the algebraic 
connection between $S^\mu$ and 
$\overline{\psi} \gamma^\mu\gamma^5 \psi$ 
allows us to express the spinor components 
$\psi_a$ $(a=1,2,3,4)$ in terms of the components of the 
torsion vector $S^\mu$ $(\mu=0,1,2,3)$ and then to 
substitute  this back into the Heisenberg equation (\ref{5}). 
In this way the Heisenberg equation 
becomes a dynamical equation for torsion 
so that {\it the torsion becomes a propagating 
field.} This leads to an essential change of the 
physical interpretation of EC gravity 
coupled to a classical spinor field. Using the 
relationship between the spinor field and the torsion
we can replace the spinor field in the Heisenberg 
equation by the torsion and view this theory as
{\it a pure vacuum 
gravitational theory with propagating torsion.} 
In the rest of this paper we will refer to such a
theory, where the torsion has become a propagating degree
of freedom via the switch with the spinor field, as
{\it Einstein-Cartan-Heisenberg gravity}. 
The Heisenberg equation (\ref{5}) now becomes 
a dynamical equation for torsion  which is of a form
similar to the non-linear, spinor equations investigated
by Ivanenko \cite{iv} and Heisenberg \cite{hz1}, \cite{hz2}.

\section{Heisenberg gravity}

We now want to examine the Einstein-Cartan-Heisenberg
gravitational theory in the limit where the metric
degrees of freedom are frozen out.
Ref.\cite{hehl1} gives a classification 
of different gravity theories with curvature and/or
torsion:
\begin{enumerate}
\item
Riemann-Cartan space $U_4$ has both curvature and 
torsion.
\item
Weitzenb\"ock space $W_4$ has only torsion; 
{\it curvature=0}.
\item
Riemann space $V_4$ has only curvature; 
{\it torsion=0}.
\item
Minkowski space $M_4$ has {\it curvature=torsion=0}.
\end{enumerate}
EC gravity is associated with $U_4$, and ordinary
Einstein gravity is associated with $V_4$. 
The gravitational theory associated
with  $W_4$, Weitzenb\"ock space, we will call 
Heisenberg gravity. With the connection between
the spinor field and the torsion the Heisenberg equation 
for the spinor field becomes the gravitational equation for 
the torsion. The implicit form for this equation is 
\begin{eqnarray}
S^{abc} = 4il_{Pl}^2\left (\overline{\psi }
\gamma ^{[a}\gamma ^b\gamma ^{c]}\psi \right ), &&
\label{8} \\
\left( i\gamma ^\mu\partial _\mu + l_{Pl}^2(\overline{\psi }
\gamma ^{[a}\gamma ^b\gamma
^{c]}\psi )\gamma _{[a}\gamma _b\gamma _{c]} - 
\frac{mc}\hbar \right) \psi 
&=&0,  
\label{9}
\end{eqnarray}

\par
Now we examine two solutions in this theory.

\subsection{4D Trivial solution}

We consider the constant spinor field in 4D Weitzenb\"ock 
spacetime where $\omega_{ab\mu} = 0$ and we further assume $m=0$ 
\footnote{As mentioned in section 2 Heisenberg's viewpoint
was that $m = 0$ was the most natural condition,
and the mass should be determined dynamically via the
time-time part of the energy-momentum density}.
:
\begin{equation}
\psi (r,t)=\left\{ 
\begin{array}{c}
a \\ 
b \\ 
c \\ 
d
\end{array}
\right\} ,  
\label{10}
\end{equation}
here $a,b,c,d$ are constants. The spin density vector 
$S^\mu \propto (\overline{\psi} \gamma^\mu\gamma^5 \psi/i)$ 
(here we use $\gamma^5 = \gamma^0\gamma^1\gamma^2\gamma^3$)  
then takes on the following form: 
\begin{eqnarray}
S^0 \propto -2 \rm{Re} (ac^* + bd^*), 
\label{11-1} \\
S^1 \propto -2 \rm{Re} (ab^* + cd^*), 
\label{11-2} \\
S^2 \propto 2 \rm{Im} (ab^* + cd^*), 
\label{11-3} \\
S^3 \propto |a|^2 - |b|^2 + |c|^2 -|d|^2
\label{11-4}
\end{eqnarray}
Combining this with the Heisenberg equation (\ref{9})
we find the following relationships between the constants 
\begin{eqnarray}
a|a|^2 + a|b|^2 - a|d|^2 -a^*c^2 + bcd^* -b^*cd = 0, 
\label{12-1} \\
|a|^2b + ac^*d - a^*cd + b^*|b|^2 - b|c|^2 - b^*d^2 = 0, 
\label{12-2} \\
a^2c^* + abd^* - ab^*d + |b|^2c - c|c|^2 - c|d|^2 = 0, 
\label{12-3} \\
|a|^2d + abc^* - a^*bc + b^2d^* - |c|^2d - d|d|^2 = 0
\label{12-4}
\end{eqnarray}
These equations have the following 
simple solution: 
\begin{equation}
a=b=c=d
\label{13-1}
\end{equation}
The simplicity of this solution is a result of
the nonlinearity of the Heisenberg equation.
This simple, constant solution is closely
related to that found in Ref. \cite{ewm3}. 
The spin density is: 
\begin{equation} 
S^\mu \propto \left \{ |a|^2; |a|^2; 0;0 \right \}.
\label{14-1}
\end{equation}
This indicates that the spin density is a constant vector 
in the $(tx)$ plane.

Another interesting aspect of this trivial solution is that
it contains no field energy-momentum in $\psi $. Using Eq. (\ref{10})
in Eq. (\ref{7}) immediately gives $T_{\mu \nu} ^D = 0$ since
$\psi$ has neither time or spatial dependence.

\subsection{5D Spherically symmetric solution}

It is straightfoward to generalize much of the 4D
development to 5D Kaluza-Klein theory. In doing this
we will work with a 5D Weitzenb\"ock spacetime ({\it i.e.}
with the metric degrees of freedom being frozen), and
we will again take the torsion to be totally antisymmetric.
In this way we find that the 5D Heisenberg equations
are practically identical to the 4D Heisenberg equations.
This allows us to take over some of the well known
4D, finite energy solutions \cite{ewm3} \cite{fin1},
and use them as solutions for the 5D Heisenberg equation. 
Varying the torsion and spinor fields 
leads to the following field equations which, except for
the range of the indices, is similar to the 4D Heisenberg
equations (\ref{5}) and (\ref{9}) 
\begin{eqnarray}
S^{abc} = 4il_{Pl}^2\left (\overline{\psi }
\gamma ^{[a}\gamma ^b\gamma ^{c]}\psi \right ), &&
\label{11} \\
\left( i\gamma ^A\partial _A + l_{Pl}^2(\overline{\psi }
\gamma ^{[a}\gamma ^b\gamma
^{c]}\psi )\gamma _{[a}\gamma _b\gamma _{c]} - 
\frac{mc}\hbar \right) \psi 
&=&0,  
\label{12}
\end{eqnarray}
here $S^{abc}$ is the 5D antisymmetric torsion tensor; 
$a,b,c = 0,1,2,3,4$ are 
five-bein indices; $A=0,1,2,3,4$ is the 5D world index; 
$\{\gamma^A, \gamma^B\} = 2 \eta^{AB}$; 
$\eta^{AB} = diag(1,-1,-1,-1,-1)$. 
The Dirac matrix $\gamma ^4$ is defined to
be $\gamma ^4 \equiv \gamma ^5$.
The ansatz for Eq. (\ref{12}) is taken as the
standard spherically symmetric spinor: 
\begin{equation}
\psi (r,t)=e^{-i\omega t}\left\{ 
\begin{array}{c}
i f(r) \\ 
0 \\ 
g(r)\cos \theta  \\ 
g(r)\sin \theta e^{i\varphi }
\end{array}
\right\} ,  
\label{13}
\end{equation}
here $r,\theta ,\varphi $ are the spherical coordinates. 
The substitution of ansatz (\ref{13}) into 
Heisenberg's Eq. (\ref{12}) gives us 
the following two equations: 
\begin{eqnarray}
g^{\prime } + (-m + \omega )f + \frac{2g}r - 
12 l_{Pl}^2f\left( f^2-g^2\right)  & = & 0,
\label{14} \\
f^{\prime } - (m + \omega )g - 12 l_{Pl}^2g
\left( f^2-g^2\right) & = & 0  
\label{15}
\end{eqnarray}
here $\hbar ,c=1$. 
These equations of the 5D Heisenberg equation
coincide identically with equations 
for the 4D Heisenberg equation, which have been 
investigated numerically in Ref. \cite{fin1} with other
nonlinear terms ({\it e.g.} $(\overline{\psi }\psi )^2$ and 
$(\overline{\psi }\gamma ^\mu\psi )^2$).
Exact solutions to the Heisenberg equation in a
curved spacetime, and with the simplified 
interaction term $(\overline{\psi }\psi )^2$,
were investigated in Ref.\cite{ewm3}. Since in the
present work we have frozen out the metric degrees
of freedom we will concentrate mainly on the
solutions given in Ref. \cite{fin1} for our
5D Kaluza-Klein system.
Equations (\ref{14}) - (\ref{15}) have regular solutions
in all space only for some discrete set of initial 
values $f_n(0)$ ($n$ is the number of 
intersections that $f(r)$ makes with the $r$-axis). 
Near the origin the regular 
solution has the following behavior: 
\begin{eqnarray}
g(r) = g_1r + g_3\frac{r^3}{6} + \cdots ,
\label{16}\\
f(r) = f_0 + f_2\frac{r^2}{2} + \cdots 
\label{17}
\end{eqnarray}
Substituting Eqs.  (\ref{16}) -(\ref{17}) into 
Eqs (\ref{14}) - (\ref{15}) gives: 
\begin{eqnarray}
g_1 = \frac{f_0}{3}\left [12l^2_{Pl} f^2_0 + (m - \omega) \right ], 
\label{18}\\
f_2 = g_1\left [12l^2_{Pl} f^2_0 + (m + \omega) \right ]. 
\label{19}
\end{eqnarray}
This means that for fixed $m$ and $\omega$ a solution depends 
only on the initial value $f(0) = f_0$.  For arbitrary $f_0$ 
values the solution is singular at infinity ($r\to \infty$). 
However there are a discrete series of $(f_0)_n$ values for which 
the solution is regular at infinity. A 
more detailed discussion of the properties of these solutions 
can be found in Ref. \cite{fin1}. 
Thus, Eqs. (\ref{14}) and (\ref{15}) have a discrete 
spectrum of regular solutions in all space, and 
they have finite energy. 
At infinity $(r\to\infty)$ these 
solutions have the following asymptotic behavior: 
\begin{eqnarray} 
f = f_\infty + \frac{ae^{-\alpha r}}{r^2} + \cdots ,
\label{20}\\
g = \frac{be^{-\alpha r}}{r^2} + \cdots ,
\label{21}\\
\alpha ^2 = 4\omega (m + \omega),
\label{22}\\
\frac{b}{a} = -\sqrt{1 - \frac{m}{\omega}}, 
\label{23}\\ 
f_\infty = \pm \sqrt{\frac{-m + \omega}{12 l^2_{Pl}}}. 
\label{24} 
\end{eqnarray}
This guarantees the finiteness of all the physical 
parameters. As an example the field energy density, $T_{00}$,
of this solution can be finite. Inserting the general
form of the spinor from Eq. (\ref{13}) into the energy-momentum
tensor of Eq. (\ref{7}) gives
\begin{equation}
\label{24a}
T_{00} ^D = \omega (f^2 + g^2)
\end{equation}
This will give a finite energy when integrated over all space
if the ansatz function $f(r) , g(r)$ fall off rapidly enough
as $r \rightarrow \infty$. From the asymptotic form of these
functions given in Eqs. (\ref{20}) - (\ref{24}), this will
occur if $f_{\infty} =0$, which in turn implies the condition
$m = \omega$. One can interpret the finite integral of
the energy density over all space as the mass of the
solution ({\it i.e.} $M = \int T_{00} ^D d^3 x$).
\par 
We can consider 5D Kaluza-Klein spacetime as the principal 
bundle over 4D Einstein spacetime. In this case (at least locally) 
5D spacetime, which is the total space of the principal bundle, is  
$M_5=M_4 \times U(1)$, 
where $M_4$ is an ordinary 4D Einstein spacetime 
(the base of the principal bundle) and 
$U(1)$ is the electromagnetic gauge group  
(the fibre of the principal bundle). 
We note that in this case the spacetime directions 
along the $M_4$ base and $U(1)$ fibre are not equivalent. 
This is very easy to understand: the points of the 
base $M_4$ are the ordinary spacetime points and the 
$M_4$ curvature can change from point to point, but 
the points on the $U(1)$ fibre are the elements of the 
gauge group and hence the curvature of the fibre is constant 
(it can depend only on spacetime points of the base). 
From a 4D observer's point of view we can introduce the 
4D spin vector $S^\mu \propto (\overline\psi\gamma^\mu\gamma^5\psi )$ 
which for the ansatz of Eq. (\ref{13}) becomes  
\begin{equation}
S^\mu \propto \left \{ 0; -g^2\sin 2\theta \cos \varphi ;
-g^2 \sin 2\theta \sin \varphi ;-f^2-g^2\cos 2\theta 
\right \}
\label{25}
\end{equation}
This is the spin density from pure vacuum gravity, 
without any matter field. In this paper, after 
freezing out the metric degrees of freedom, we are left
with only torsion. This torsion is interpreted as
the source of the classical, auxiliary spinor field
$\psi$, in both 4D and 5D versions of the theory. This is an
inversion of the usual interpretation where the spinor
is considered as fundamental and acts as the source for
the torsion. 
\par 
In Ref. \cite{kurd} solutions for the nonlinear Heisenberg 
equation were obtained which in contrast to our localized 
nonsingular solution have the form of elliptic functions.

\section{Physical discussion}

In this paper we considered 4D and 5D Heisenberg gravity,
where the only dynamical,  
gravitational degree of freedom was the torsion. 
Usually torsion is a non-propagating, non-dynamical
degree of freedom. Here the dynamical behavior of the
torsion is defined by the nonlinear Heisenberg 
equation (Eqs. (\ref{5}) , (\ref{9}) or (\ref{12}))
through an auxiliary classical spinor field 
(it is possible that this field has some independent 
physical meaning but this is a problem for a
future investigation). A nice feature of this
dynamical torsion theory is that it possesses
some simple, non-singular solutions ({\it e.g.}
Eqs. (\ref{20}) - (\ref{24})) which may yield
a physical geometric model for spin. In contrast for
Einstein gravity, where the torsion $=0$, the spherically
symmetric Schwarzschild solution, while thought
to give a good description of certain physical
situations such as supermassive collapsed stars,
nevertheless possesses unphysical singularities.  
This absence of singularities for these solutions
of Heisenberg gravity may imply that the full
Einstein-Cartan-Heisenberg gravity may also have
singularity free solutions as compared to
ordinary Einstein gravity. The main point
that we want to emphasize however is that 
Heisenberg gravity appears to give a purely gravitational
model for spin (the torsion is the source of the  auxiliary 
classical spinor field). 
\par 
Gravity can be formulated in terms
of connections and metrics, with the torsion as part
of the connection. In this paper we restricted ourselves
to only dealing with the torsion by considering
Weitzenb{\"o}ck space where $curvature =0$ so that the
metric degrees of freedom are frozen out.   
Taking into account that the torsion is related to
a bilinear combination of an auxiliary spinor field
would then lead to a quantized version of the
torsion via the quantization of the quadratic form
of the auxiliary spinor field. The apparent problem
with this is that the original Heisenberg system is hard
to quantize by conventional means. In fact because
of the non-linear fermion interaction term this
theory is conventionally non-renormalizable.
However, in  the 50's Heisenberg
\cite{hz1}, \cite{hz2} proposed a nonperturbative
method of quantizing the nonlinear Heisenberg equation.
By applying this nonperturbative method, outlined
in Ref. \cite{hz1} \cite{hz2}, to the auxiliary spinor field
it is hoped that the quantization of torsion (=part of gravity)
can be achieved through its connection to the auxiliary
spinor field. Even if such a program of quantizing
the torsion part of gravity is possible, this still
leaves open the much harder and deeper question of
how to quantize in Riemann or Riemann-Cartan space.

Another possible problem related to the quantization
of the essentially classical theory outlined here,
is that quantum effects could alter the connection
between the torsion and the auxiliary spinor field
given in Eq. (\ref{4}) (or more directly quantum
effects the could alter the connection given by 
$S_{\mu} \propto {\overline \psi} \gamma ^{\mu}
\gamma ^5 \psi$). In particular one may worry
that this relationship could be effected by the
axial anomaly \cite{ob}. Since the axial anomaly
in this context would be connected with the metric degrees
of freedom, the problem does not become an issue
in our simplified model where these degrees of freedom
have been frozen out. This possible problem would
become important once one moved from Weitzenb{\"o}ck
space to Riemann-Cartan space.
\par 
The ideas presented here have their origins in the
attempts by Wheeler, Einstein and others to geometrize 
physics. As an example one can point to Wheeler's  
geometrical model of electrical charge as a
wormhole threaded by electric flux (This ``charge without
charge'' idea can be found in Ref. \cite{wh}). 
But this model can not be a realistic model for
a charged fermion such as the electron 
since it does not have a spin angular momentum.
Wheeler wrote in Ref. \cite{wh}: 
``$\dots$ It is impossible to accept any description 
of elementary particles that does not have a place for spin 
$\frac{1}{2}$. $\dots$ Unless and until an answer is 
forthcoming, {\it pure quantum geometrodynamics 
must be judged deficient as a basis for elementary particle 
physics} $\dots$ .'' As outlined in this paper it is possible 
that Heisenberg gravity can give such a geometrical 
model of spin. In this case the spherically 
symmetric solution of Eqs. (\ref{20}) - (\ref{24})
would represent a
{\it pure geometrical model of the spin}.
In Ref.\cite{dzh2} a composite
wormhole model of electrical charge was advanced in spirit of 
Wheeler idea of ``charge without charge''. Here 
by examining a gravity theory with torsion and the 
metric degrees of freedom frozen out (Heisenberg gravity)
we arrive at a geometrical model for spin. 
Curiously in both the model of electric charge of
Ref. \cite{dzh2} and the model for spin of the
present paper it appears that a higher dimensional
manifold ({\it i.e.} 5D) is required. The eventual 
hope is that in the full 
Einstein-Cartan-Heisenberg gravity one may find 
{\it a wormhole with spin and charge} 
(``charge without charge'', ``mass without mass'' and 
``spin without spin''). Such a solution for the 
full ECH gravity would support the supposition of Einstein 
and Wheeler that {\it Nature consists from nothing?} 

\section{Acknowledgments}
VD is grateful to Georg Forster Research Fellowship 
of Alexander von Humboldt Foundation for financial 
support of this project. 

\appendix

\section{Riemann-Cartan geometry}

Here we give a simple introduction to Riemann-Cartan 
geometry following Ref. \cite{hehl}. The affine connection of 
Riemann-Cartan spacetime is: 
\begin{equation}
\Gamma^A_{\bullet BC} =  \left \{ \,^B_{AC}\right \}
+ {S_{BC}}^A_\bullet - {S_C}^A_{\bullet B} + S^A_{\bullet AB},  
\label{A12}
\end{equation}
here $\left \{ \,^B_{AC}\right \}$ are Christoffel symbols. 
Cartan's torsion tensor ${S_{BC}}^A_.$ is defined 
according to : 
\begin{equation}
{S_{BC}}^A_\bullet = S^A_{\bullet BC} = 
\frac{1}{2}\Gamma ^A_{[BC]} = 
\frac{1}{2} \left (\Gamma^A_{\bullet BC} - \Gamma^A_{\bullet CB} \right ). 
\label{A13}
\end{equation}
The contorsion tensor is: 
\begin{equation}
{K_{BC}}^A_\bullet = {S_{BC}}^A_\bullet + {S_C}^A_{\bullet B} - 
S^A_{\bullet AB}
\label{A14}
\end{equation}
In this case the affine connection is: 
\begin{equation}
\Gamma^A_{\bullet BC} = \left \{ \,^B_{AC}\right \} 
- {K_{BC}}^A_\bullet .
\label{A15}
\end{equation}
The Riemann curvature tensor is defined in the usual 
way as: 
\begin{eqnarray} 
R^A_{\bullet BCD} = \partial_C\Gamma ^A_{\bullet BD} - 
\partial_D\Gamma ^A_{\bullet BC} + 
\Gamma ^A_{\bullet EC}\Gamma ^E_{\bullet BD} - 
\Gamma ^A_{\bullet ED}\Gamma ^E_{\bullet BC} = 
\nonumber \\
\stackrel{\{\}}{R^A}_{\bullet BCD} + \nabla _D K^A_{\bullet BC} - 
\nabla _C K^A_{\bullet BD} + K^A_{\bullet EC}K^E_{\bullet BD} - 
K^A_{\bullet ED}K^E_{\bullet BC}
\label{A16}
\end{eqnarray}
A modified torsion tensor is: 
\begin{equation}
{T_{BC}}^A_\bullet = {S_{BC}}^A_\bullet + 
\delta ^A_B {S_{CD}}^D_\bullet - 
\delta ^A_C {S_{BD}}^D_\bullet .
\label{A17}
\end{equation}
We can decompose the curvature scalar into Riemannian 
and contorsion pieces as follows: 
\begin{equation}
R = \stackrel{\{\}}{R} +  
2\stackrel{\{\}}{\nabla}_A \left ({K_B^\bullet }^{AB} \right )- 
{T_A^\bullet }^{BC}{K_{CB}}^A_\bullet 
\label{A18}
\end{equation}
For antisymmetric torsion we can write: 
\begin{eqnarray}
S_{ABC} = T_{ABC}; \qquad K_{ABC} = - S_{ABC},
\label{A19-1}\\
R = \stackrel{\{\}}{R} - S_{ABC} S^{ABC}.
\label{A19-2}
\end{eqnarray}


\begin{thebibliography}{20}

\bibitem{hehl1} 
F.W.Hehl, J.Dermott McCrea, E.W. Mielke, Y. Neeman, 
Phys. Rept., {\bf 258}, 1 (1995). 

\bibitem{hehl2} 
F.W.Hehl, P. von der Heyde and G.D. Kerlick, Rev. Mod. Phys., 
{\bf 48}, 393 (1976). 

\bibitem{ewm1} E.W. Mielke, {\it Geometrodynamics of Gauge Fields} -
On the geometry of Yang-Mills and gravitational gauge theories,
(Akademie-Verlag, Berlin 1987)

\bibitem{dzh1} 
V.Dzhunushaliev, Int. J. Mod. Phys., {\bf D7}, 909 (1998). 

\bibitem{wh} 
J.A.Wheeler, {\it Geometrodynamics}, (Academic Press 
New-York and London, 1962).

\bibitem{kuc} K. Kuchar, Acta. Phys.  Pol. {\bf 28}, 695 (1965)

\bibitem{ewm2} E.W. Mielke, {\it et. al.}, ``Yang-Mills-Clifford form
of the Einstein action'', in : {\it Gravity, Particles and Space-time},
ed. P. Pronin and G. Sardanashvily (World Scientific, Singapore, 1996)
p. 217-254

\bibitem{sab}  
V. de Sabbata and C. Sivaram, {\it Spin and Torsion in 
Gravitation}, (World Scientific Publishing, 1994). 

\bibitem{hehl} 
F.W.Hehl and B.K.Datta, J.Math. Phys., {\bf 12}, 
1334 (1971). 

\bibitem{iv}
D.Ivanenko, Phys. Z. Sowjetunion, {\bf 13}, 141 (1938). 

\bibitem{hz1}
W. Heisenberg, Nachr. Akad. Wiss. G{\"o}ttingen, N8, 111 (1953);
W. Heisenberg, Z. Naturforsch., {\bf 9a}, 292 (1954); W. Heisenberg,
F. Kortel and H. M{\"u}tter, Z. Naturforsch., {\bf 10a}, 425 (1955);
W. Heisenberg, Z. f{\"u}r Phys., {\bf 144}, 1 (1956); P. Askali and
W. Heisenberg, Z. Naturforsc., {\bf 12a}, 177 (1957); W. Heisenberg,
Nucl. Phys., {\bf 4}, 532 (1957); W. Heisenberg, Rev. Mod. Phys., 
{\bf 29}, 269 (1957)

\bibitem{hz2} 
W.Heisenberg, {\it Introduction to the unified
field theory of elementary particles.}, (Wiley, London, 1966). 

\bibitem{ewm3} E.W. Mielke, J. Math. Phys., {\bf 22}, 2034 (1981)

\bibitem{fin1}  
R. Finkelstein, R.Lelevier and M. Ruderman, 
Phys. Rev., {\bf 83}, 326 (1951); 
R. Finkelstein, C. Fronsdal and P.Kaus, 
Phys. Rev., {\bf 103}, 1571 (1956). 

\bibitem{kurd} 
D.Kurdgelaidze, JETP, {\bf 32}, 1156 (1957); {\bf 34}, 
1587 (1958); {\bf 36}, 842 (1959) (in Russian). 

\bibitem{ob} Y. Obukhov {\it et. al.}, Found. Phys. {\bf 27},
1221 (1997); E.W. Mielke and D. Kreimer, Int. J. Mod. Phys.
{\bf D7}, 535 (1998)

\bibitem{dzh2} 
V.Dzhunushaliev, ``Multidimensional geometrical model 
of the renormalized electrical charge with splitting off 
the extra coordinates'', awarded  {\it Honorable Mention}
by Grav. Res. Found., 1998; 
Mod. Phys. Lett.A, {\bf 13}, 2179 (1998).

\end{thebibliography}
\end{document}